\documentclass{ws-procs9x6}

\usepackage{amsfonts}
\usepackage{amssymb}
\usepackage{amsmath}

\usepackage{cite}
\usepackage{epsf}

\def\nd{\end{document}}

\def\tchi{{\tilde\chi}}

\def\hh{h}

\def\nn{\nonumber}
\def\nt{\noindent}

\def\han{{\textstyle\frac{p+q-2}{2}}}

  \def\tV{{\tilde V}}
\def\tcn{{\tilde{\cal N}}}

\def\bu{\noindent $\bullet~$}

\def\deg{{\rm deg}\,}

\def\riga{-\kern-4pt - \kern-4pt -}
\font\fat=cmsy10 scaled\magstep5

\def\Bbullet{\raise-3pt\hbox{\fat\char"0F}}

\def\down{\raise1.5pt\hbox{$\phantom{a}_2$}\downarrow}

\def\downa{\raise1.5pt\hbox{$\phantom{a}_{2\atop m_2}$}\downarrow}

\def\llr{\longrightarrow}

\def\({\left(}
\def\){\right)}
\def\eps{\epsilon}
 
\def\lra{\longrightarrow}

\def\ha{{\textstyle{\frac{1}{2}}}}

\def\bbz{\mathbb{Z}}
\def\bbc{\mathbb{C}}
\def\bac{\bbc} 
\def\bbn{\mathbb{N}}

\def\b{\beta}

\def\vr{\vert}

\def\ed{\end{document}}

\def\ca{{\cal A}}  \def\cc{{\cal C}}
\def\cd{{\cal D}}  \def\cf{{\cal F}}
\def\cg{{\cal G}} \def\ch{{\cal H}} 
  
\def\cm{{\cal M}} \def\cn{{\cal N}} 
\def\cp{{\cal P}}  
 \def\ct{{\cal T}}

\def\ido{intertwining differential operator}
\def\idos{intertwining differential operators}

\def\L{\Lambda}
\def\r{\rho}

\begin{document}

\title{Conservation Laws for SO(p,q)}

 \author{V.K. Dobrev}

 \address{Institute for Nuclear Research and Nuclear Energy, Bulgarian Academy of Sciences,
 Tsarigradsko Chaussee 72, BG-1784 Sofia, Bulgaria}

\begin{abstract}
Using our previous results on the systematic construction of
invariant differential operators for non-compact semisimple Lie
groups we classify the conservation laws in the case of SO(p,q).
\end{abstract}

\bodymatter

\section{Introduction}

In a recent paper \cite{Dobinv} we started the systematic explicit construction
of invariant differential operators. We gave an explicit description
of the building blocks, namely, the {\it parabolic subgroups and
subalgebras} from which the necessary representations are induced.
Thus we have set the stage for study of different non-compact
groups.

Since the study and description of detailed classification should be
done group by group we had to decide which groups to study first. We
decided to start with a subclass of   the hermitian-symmetric
algebras which share some special properties of the conformal
algebra ~$so(n,2)$. That is why, in view of applications to physics,
we called these algebras
  '{\it conformal Lie algebras}' (CLA), (or groups)
  \cite{Dobeseven}{}.
Later we gave a natural way to go beyond this subclass using essentially
the same results. For this we introduce the new notion of ~{\it
parabolic relation}~ between two non-compact semisimple Lie algebras
$\cg$ and $\cg'$ that have the same complexification and possess
maximal parabolic subalgebras with the same complexification
\cite{Dobparel}{}.

Thus, for example, using results for the conformal algebra
~$so(n,2)$~ (for fixed $n$) we can obtain results for all
pseudo-orthogonal algebras ~$so(p,q)$~ such that ~$p+q=n+2$. In this
way, in \cite{Dobparel} (among other things)  we gave the main and
the reduced multiplets of indecomposable elementary representations
for $so(p,q)$ including the necessary data for all relevant
invariant differential operators. We specially stressed that the
classification of all invariant differential operators includes as
special cases all possible ~{\it conservation laws}~ and ~{\it
conserved currents}, unitary or not. In the present paper we give
explicitly the conservation laws in the case of ~$so(p,q)$.

This paper is a short sequel of \cite{Dobparel}{}, based on Invited
talk at Group-29, Chern Institute of Mathematics, Nankai U.,
20-26.8.2012.  Due to the lack of space we refer to \cite{Dobparel}
for motivations and extensive list of literature on the subject.

\section{Preliminaries}

\nt Let ~$\cg=so(p,q)$, ~$p\geq q$, ~$p+q>4$.
 We choose a maximal parabolic $\cp = \cm \oplus \ca \oplus \cn$
 such that:
\begin{equation}\label{cmsosmay} \cm ~=~  so(p-1,q-1),\quad   \dim\, \ca ~=~1 \  ,\quad   \dim\, \cn ~=~   p+q
-2 \ .\end{equation} With this choice we get for the conformal algebra $so(n,2)$
the Bruhat decomposition ~$\cg = \cp \oplus \tcn$~ with direct  physical meaning
($\tcn\cong \cn$) \cite{Dobparel}{}.

 We label   the signature of the representations of
$\cg$   as follows: \begin {eqnarray}\label{sgnd}  &&\chi ~~=~~ \{\,
n_1\,, \ldots,\, n_{\hh}\,;\, c\, \} \ , \\  &&\quad n_j \in \bbz/2\
, \quad c=d-\han\ , \quad \hh \equiv [\han] , \nn\\  && \vr n_1 \vr
< n_2 < \cdots <  n_{\hh}\ , \quad p+q ~~{\rm even}\ , \nn\\  && 0 <
n_1 < n_2 < \cdots <  n_{\hh} \ , \quad p+q ~~{\rm odd}\ ,
\nn\end{eqnarray} where the parameter $c$ (related to the conformal weight $d$) labels the
characters of $\ca$,  and the first $\hh$ entries are labels of the
finite-dimensional (nonunitary for $q\neq 1$) irreps ~$\mu$~ of ~$\cm\,$.

 We call the above induced
representations ~$\chi =$ Ind$^G_{\cp}(\mu\otimes\nu \otimes 1)$~
~{\it  elementary representations} of $G=SO(p,q)$ \cite{DMPPT}{}.
 Their spaces of functions are:  \begin {eqnarray}\label{func}
\cc_\chi ~&=&~ \{ \cf \in C^\infty(G,V_\mu) ~ \vr ~ \cf (gman) ~=~
e^{-\nu(H)} \cdot D^\mu(m^{-1})\, \cf (g) \} \nn\end{eqnarray}  where ~$a=
\exp(H)$, ~$H\in\ca\,$, ~$m\in M=SO(p-1,q-1)$, ~$n\in N=\exp\cn$. The
representation action is the {\it left regular action}:  \begin{equation}\label{lrega}
(\ct^\chi(g)\cf) (g') ~=~ \cf (g^{-1}g') ~, \quad g,g'\in G\ .\end{equation}

\bu An important ingredient in our considerations are the ~{\it \it
highest/lowest weight representations}~ of ~$\cg^\bac$. These can be
realized as (factor-modules of) Verma modules ~$V^\L$~ over
~$\cg^\bac$, where ~$\L\in (\ch^\bac)^*$, ~$\ch^\bac$ is a Cartan
subalgebra of ~$\cg^\bac$, weight ~$\L = \L(\chi)$~ is determined
uniquely from $\chi$ \cite{Dob}{}.

Actually, since our ERs are induced from finite-dimensional
representations of ~$\cm$~   the Verma modules are
always reducible. Thus, it is more convenient to use ~{\it
generalized Verma modules} ~$\tV^\L$~ such that the role of the
highest/lowest weight vector $v_0$ is taken by the
(finite-dimensional) space ~$V_\mu\,v_0\,$. For the generalized
Verma modules (GVMs) the reducibility is controlled only by the
value of the conformal weight $d$, or the parameter $c$. Relatedly, for the \idos{} only
the reducibility w.r.t. non-compact roots is essential.

\bu One main ingredient of our approach is as follows. We group the
(reducible) ERs with the same Casimirs in sets called ~{\it
multiplets} \cite{Dobmul}{}. The multiplet corresponding to fixed values of the
Casimirs may be depicted as a connected graph, the {\it vertices} of which
correspond to the reducible ERs and the {\it lines (arrows)}  between the vertices
correspond to intertwining operators.
The multiplets contain explicitly all the data necessary to
construct the \idos{}. Actually, the data for each \ido{} consists
of the pair ~$(\b,m)$, where $\b$ is a (non-compact) positive root
of ~$\cg^\bac$, ~$m\in\bbn$, such that the {\it BGG  Verma module
reducibility condition} \cite{BGG} (for highest weight modules) is fulfilled:
\begin{equation}\label{bggr} (\L+\r, \b^\vee ) ~=~ m \ , \quad \b^\vee \equiv 2 \b
/(\b,\b) \ \end{equation} where $\r$ is half the sum of the positive roots of
~$\cg^\bac$. When the above holds then the Verma module with shifted
weight ~$V^{\L-m\b}$ (or ~$\tV^{\L-m\b}$ ~ for GVM and $\b$
non-compact) is embedded in the Verma module ~$V^{\L}$ (or
~$\tV^{\L}$). This embedding is realized by a singular vector
~$v_s$~  expressed by a polynomial ~$\cp_{m,\b}(\cg^-)$~ in the universal
enveloping algebra ~$(U(\cg_-))\ v_0\,$, ~$\cg^-$~ is the subalgebra
of ~$\cg^\bac$ generated by the negative root generators \cite{Dix}{}.
 More explicitly, \cite{Dob} ~$v^s_{m,\b} = \cp_{m,\b}\, v_0$ (or ~$v^s_{m,\b} =
 \cp_{m,\b}\, V_\mu\,v_0$ for GVMs).\footnote{For
explicit expressions for singular vectors we refer to \cite{Dobsin}{}.}\\
   Then there exists \cite{Dob} an ~{\it \ido{}}~ of order ~$m=m_\b$~:
\begin{equation}\label{invop}
  \cd_{m,\b} ~:~ \cc_{\chi(\L)}
~\llr ~ \cc_{\chi(\L-m\b)} \end{equation} given explicitly by: \begin{equation}\label{singvv}
 \cd_{m,\b} ~=~ \cp_{m,\b}(\widehat{\cg^-})  \end{equation} where
~$\widehat{\cg^-}$~ denotes the {\it right action} on the functions
~$\cf\,$.

Thus, in each such situation we have an ~{\it invariant differential equation}~ of order ~$m=m_\b$~:
\begin{equation}\label{invde} \cd_{m,\b}\ f ~=~ f' \ , \qquad f \in \cc_{\chi(\L)} \ , \quad
f' \in \cc_{\chi(\L-m\b)} \ .\end{equation}

In most of these situations the invariant operator ~$\cd_{m,\b}$~ has a non-trivial invariant
kernel in which a subrepresentation of $\cg$ is realized. Thus, studying the equations
with trivial RHS:
\begin{equation}\label{invdec} \cd_{m,\b}\ f ~=~ 0 \ , \qquad f \in \cc_{\chi(\L)} \ ,
   \end{equation}
is also very important. For example, in many physical applications
 in the case of first order differential operators,
i.e., for ~$m=m_\b = 1$, equations \eqref{invdec}
are called ~{\it conservation laws}, and the elements ~$f\in \ker \cd_{m,\b}$~
are called ~{\it conserved currents}.  Below we give them explicitly for ~$so(p,q)$.

\section{Classification of  Conservation Laws for  ~{\boldmath $so(p,q)$}}

Using results of \cite{DMPPT,DoPe:78,PeSo,Dobsrni,Dobpeds} we
present the main multiplets (which contain the maximal number of ERs
with this parabolic) with the following explicit parametrization of
the ERs in the multiplets (following \cite{Dobsrni}): \begin
{eqnarray}\label{sgne} \chi^\pm_1 &=& \{ \eps\, n_1\,, \ldots,\,
n_\hh \,;\, \pm n_{\hh+1} \} \ ,
 \quad n_\hh < n_{\hh+1}\ , \nn\\
\chi^\pm_2 &=& \{ \eps\,  n_1\,, \ldots,\, n_{\hh-1}\,,\,
n_{\hh+1}\,;\, \pm n_\hh \}     \nn\\  \chi^\pm_3  &=&  \{  \eps\,
n_1,  \ldots,  n_{\hh-2},  n_{\hh},  n_{\hh+1}\,;\, \pm
n_{\hh-1} \}     \nn\\  ... \\
 \chi^\pm_{\hh-1} &=& \{ \eps\, n_1\,,
n_2\,, n_4\,,\ldots,\, n_{\hh}\,,\, n_{\hh+1}\,;\, \pm  n_3 \}  \nn\\
 \chi^\pm_{\hh} &=& \{ \eps\, n_1\,,
n_3\,, \ldots,\, n_{\hh}\,,\, n_{\hh+1}\,;\, \pm  n_2 \}     \nn\\
\chi^\pm_{\hh+1} &=& \{ \eps\, n_2\,, n_3\,, \ldots,\, n_{\hh}\,,\,
n_{\hh+1}\,;\, \pm  n_1 \}     \nn\\  &\eps =& \begin{cases}
 \pm\,, ~&~  p+q ~~ even  \nn\\
                     1,  ~&~  p+q  ~~ odd \end{cases} \nn\end{eqnarray}
($\eps = \pm$~ is correlated with $\chi^\pm$).
Clearly, the multiplets correspond 1-to-1 to the finite-dimensional irreps
of $so(p+q,\bbc)$ with signature $\{n_1,\ldots,n_h,n_{h+1}\}$ and we are able to use
previous results due to the parabolic relation between the $so(p,q)$ algebras
for $p+q$~-fixed.
Note that the two representations in each pair ~$\chi^\pm$~ are called ~{\it shadow fields}.

 Further, we denote by ~$\cc^\pm_i$~ the representation space with signature
~$\chi^\pm_i\,$.

The ERs in the multiplet are related by {\it intertwining integral and
differential operators}.

The  {\it integral operators} were introduced by
Knapp and Stein \cite{KnSt}{}.
Here these operators intertwine the pairs ~$\cc^\pm_i$~ (cf.
\eqref{sgne}): \begin{equation}\label{knapps} G^\pm_i ~:~ \cc^\mp_i \lra \cc^\pm_{i}
\ , \quad i ~=~ 1,\ldots,1+\hh  \ . \end{equation}

The {\it \idos}\ correspond to non-com\-pact positive roots of the
root system of ~$so(p+q,\bbc)$, cf. \cite{Dob}{}. In the current
context, compact roots of $so(p+q,\bbc)$ are those that are roots
also of the subalgebra $\cm^\bbc$, the rest of the roots are
non-compact. We denote the differential operators by
~$d_i\,,d'_i\,$. The spaces from \eqref{sgne} they intertwine   are:
\begin {eqnarray}\label{idoseven} && d_i ~:~ \cc^-_i ~\lra~ \cc^-_{i+1} \ , \quad i = 1,\ldots,h ~ ;\nn\\
&& d'_i ~:~ \cc^+_{i+1} ~\lra~ \cc^+_{i} \ , \quad i = 1,\ldots,h-1 ~ ;\nn\\
&& d_h ~:~ \cc^+_{h+1} \lra \cc^+_{h} \ , \quad (p+q)-{\rm even}~; \nn\\
&& d'_h ~:~ \cc^-_{h} \lra \cc^+_{h+1} \ , \quad (p+q)-{\rm even}~; \nn\\
&& d'_h ~:~ \cc^-_{h+1} \lra \cc^+_{h} \ , \quad (p+q)-{\rm even}~; \nn\\
&& d'_{h} ~:~ \cc^+_{h+1} \lra \cc^+_{h} \ , \quad (p+q)-{\rm odd}~; \nn\\
&& d_{h+1} ~:~ \cc^-_{h+1} \lra \cc^+_{h+1} \ , \quad (p+q)-{\rm odd}~ .
 \end{eqnarray}
The degrees of these \idos\ are given just by the differences of the
~$c$~ entries \cite{Dobsrni}{}: \begin {eqnarray}\label{degr} &&\deg
d_i = \deg d'_i = n_{\hh+2-i} - n_{\hh+1-i} \,, \qquad i =
1,\ldots,h \,,
\\  &&
\deg d'_{h} = n_2 + n_1
\,, \quad (p+q) - {\rm even}\ , \nn\\  &&
\deg d_{h+1} = 2n_1 \,, \quad (p+q) - {\rm odd} \ .
\nn\end{eqnarray}
where $d'_h$ is omitted from the first line for $(p+q)$ even.

Thus, we are able to list all cases of ~{\bf first order} \idos\ or ~{\bf conservation laws}
 (we give only the signatures):
\begin {eqnarray}\label{idoford}
 && d_1 ~:~ \{
\eps\, n_1\,,\,n_2\,, \ldots,\, n_\hh \,;\, - n_{\hh}-1 \}  ~\lra~
\{ \eps\, n_1\,,\,n_2\,, \ldots,\,n_{h-1}\,, n_\hh+1 \,;\, -
n_{\hh} \}\ ; \nn\\ && d'_1 ~:~ \{ n_1\,,\ldots,\,
n_{h-1}\,,n_\hh+1 \,;\,  n_{\hh} \} ~\lra~
\{ n_1\,, \ldots,\, n_\hh \,;\,  n_{\hh}+1 \}  \ ; \nn\\
&& \ldots  \\
&& d_i ~:~ \{\eps\, n_1\,,\,n_2\,,\ldots,\,n_{\hh+1-i}\,,n_{\hh+3-i}\,, \ldots\,,
 n_{\hh+1} \,;\, - n_{\hh+1-i}-1 \}  ~\lra~ \nn\\   &&\qquad\quad
\{ \eps\, n_1\,,\,n_2\,, \ldots,\, n_{\hh-i}\,, n_{\hh+1-i}+1\,,n_{\hh+3-i}\,, \ldots\,,
n_{\hh+1} \,;\, -  n_{\hh+1-i} \}\ , \nn\\   &&\qquad\quad\quad i = 2,\ldots,h-1 ~ ; \nn\\
&& d'_i ~:~ \{  n_1\,,\, \ldots,\, n_{\hh-i}\,, n_{\hh+1-i}+1\,,n_{\hh+3-i}\,, \ldots\,,
n_{\hh+1} \,;\,   n_{\hh+1-i} \} ~\lra~ \nn\\   &&\qquad\quad
\{ n_1\,,\,\ldots,\,n_{\hh+1-i}\,,n_{\hh+3-i}\,, \ldots\,,
 n_{\hh+1} \,;\,  n_{\hh+1-i}+1 \} \ , \nn\\   &&\qquad\quad\quad i = 2,\ldots,h-1 ~ ; \nn\\
&& \ldots  \nn\\
&& d_h ~:~ \{\eps\, (n_2-1)\,,\,n_{3}\,, \ldots\,,
 n_{\hh+1} \,;\, - n_{2} \}  ~\lra~ \nn\\   &&\qquad\quad
\{ \eps n_2\,,n_{3}\,, \ldots\,,
n_{\hh+1} \,;\, 1-  n_{2} \}\ , \quad n_2>\ha\ ; \nn\\
&& d_h ~:~ \{   n_{2}\,,n_{3}\,, \ldots\,,
n_{\hh+1} \,;\,   n_{2}-1 \} ~\lra~
\{ n_2-1\,,\,n_{3}\,, \ldots\,,
 n_{\hh+1} \,;\,  n_{2} \}
\ ,  \nn\\   &&\qquad\quad\quad\quad n_2>\ha\ , \quad (p+q) - {\rm even}; \nn\\
&& d'_h ~:~ \{  n_2-1\,,\,n_{3}\,, \ldots\,,
 n_{\hh+1} \,;\,  -n_{2} \}  ~\lra~
\{   n_{2}\,,n_{3}\,, \ldots\,,
n_{\hh+1} \,;\,  1- n_{2} \} \ , \nn\\   &&
\qquad\quad\quad\quad n_2>\ha\ , \quad (p+q) - {\rm even}\ ; \nn\end{eqnarray}
\begin {eqnarray} && d'_h ~:~ \{  -n_2\,,\,n_{3}\,, \ldots\,,
 n_{\hh+1} \,;\,  n_{2}-1 \}  ~\lra~
\{  1- n_{2}\,,n_{3}\,, \ldots\,,
n_{\hh+1} \,;\,  n_{2} \} \ , \nn\\   &&
\qquad\quad\quad \quad n_2>\ha\ , \quad (p+q) - {\rm even}\ ; \nn\\
&& d'_h ~:~ \{  n_2\,,\,n_{3}\,, \ldots\,,
 n_{\hh+1} \,;\,  n_{2}-1 \}  ~\lra~
\{   n_{2}-1\,,n_{3}\,, \ldots\,,
n_{\hh+1} \,;\,   n_{2} \} \ , \nn\\   &&
\qquad\quad\quad \quad n_2>1\ , \quad (p+q) - {\rm odd}\ ; \nn\\   &&
d'_{h+1} ~:~ \{  -n_2\,,\,n_{3}\,, \ldots\,,
 n_{\hh+1} \,;\,  -\ha \}  ~\lra~
\{   n_{2}\,,n_{3}\,, \ldots\,,
n_{\hh+1} \,;\,   \ha \} \ , \nn\\   &&
\qquad\quad\quad\quad n_2>1\ , \quad (p+q) - {\rm odd}\ ; \nn\\   &&
\qquad\qquad \eps = - (-1)^{p+q} \ .\nn \end{eqnarray} Some of these
conservation laws were given also in \cite{Mets}{}. Although the
operators  are valid for arbitrary  $so(p,q)$ ($p+q\geq 5$) the
contents of the ERs is very different. This analysis was done in
detail in \cite{Dobparel}{}.

Besides the above cases there are other ERs (not related to finite-dimensional irreps of
$so(p+q,\bbc)$), and which also give rise to first order \idos, resp. conservation laws:
\begin {eqnarray}\label{sgner} && d_\eps ~:~ \tchi^-_\eps ~\lra~ \tchi^+_\eps \ ,  \\
 &&\tchi^\pm_{\eps} = \{ \pm\eps\, \ha\,,
n_3\,, \ldots,\,   n_{\hh+1}\,;\, \pm  \ha \} \,, ~ \nn
\end{eqnarray}
where ~$\eps$~ is defined as in \eqref{sgne}, but here it is not correlated with $\tchi^\pm\,$.
  There is also    a Knapp-Stein integral operator
acting from ~$\tchi^+_\eps\,$ to ~$\tchi^-_\eps\,$.
The representations  ~$\tchi^-_\eps$~ are called ~{\it minimal representations}.

\end{document}